\documentclass[12pt]{article}

\usepackage{enumitem}
\usepackage{graphicx}
\usepackage{float}
\usepackage{cite}
\usepackage{amsfonts}
\usepackage{amssymb}
\usepackage{amsmath}
\usepackage{xcolor}

\def\comp{{\rm C}\llap{\vrule height7.1pt width1pt depth-.4pt\phantom t}}
\def\gtwid{\mathrel{\raise.3ex\hbox{$>$\kern-.75em\lower1ex\hbox{$\sim$}}}}
\def\ltwid{\mathrel{\raise.3ex\hbox{$<$\kern-.75em\lower1ex\hbox{$\sim$}}}}
\def\square{\kern1pt\vbox{\hrule height 1.2pt\hbox{\vrule width 1.2pt\hskip 3pt
  \vbox{\vskip 6pt}\hskip 3pt\vrule width 0.6pt}\hrule height 0.6pt}\kern1pt}

\begin{document}

\begin{titlepage}

\begin{flushright}
CCTP-2024-7 \\
UFIFT-QG-24-04
\end{flushright}

\vskip 1cm

\begin{center}
{\bf Summing Gravitational Effects from Loops of Inflationary Scalars}
\end{center}

\vskip 1cm

\begin{center}
S. P. Miao$^{1\star}$, N. C. Tsamis$^{2\dagger}$ and 
R. P. Woodard$^{3\ddagger}$ (Corresponding author)
\end{center}

\vskip 0.5cm

\begin{center}
\it{$^{1}$ Department of Physics, National Cheng Kung University, \\
No. 1 University Road, Tainan City 70101, TAIWAN}
\end{center}

\begin{center}
\it{$^{2}$ Institute of Theoretical Physics \& Computational Physics, \\
Department of Physics, University of Crete, \\
GR-710 03 Heraklion, HELLAS}
\end{center}

\begin{center}
\it{$^{3}$ Department of Physics, University of Florida,\\
Gainesville, FL 32611, UNITED STATES}
\end{center}

\vspace{0cm}

\begin{center}
ABSTRACT
\end{center}
We develop a procedure for re-summing the large logarithms induced
in gravity by loops of inflationary scalars. We first show how the
scalar can be integrated out of the field equations in the presence
of constant graviton field. We then extend this result to a fully
conserved form which explains the need for a finite renormalization
of the cosmological constant which was previously inferred from 
explicit computation. A variant of the renormalization group turns
out to explain the large logarithmic corrections revealed by 
explicit computation in the electric field strength of gravitational 
radiation and in the potentials which characterize the response to
a point mass. The implications for graviton loops are discussed. 

\begin{flushleft}
PACS numbers: 04.50.Kd, 95.35.+d, 98.62.-g
\end{flushleft}

\vskip 0.5cm

\begin{flushleft}
$^{\star}$ e-mail: spmiao5@mail.ncku.edu.tw \\
$^{\dagger}$ e-mail: tsamis@physics.uoc.gr \\
$^{\ddagger}$ e-mail: woodard@phys.ufl.edu
\end{flushleft}

\end{titlepage}

\section{Prologue}

The geometry of cosmology can be characterized by a scale factor $a(t)$,
Hubble parameter $H(t)$ and first slow roll parameter $\epsilon(t)$,
\begin{equation}
ds^2 = - dt^2 + a^2(t) d\vec{x} \!\cdot\! d\vec{x}
\qquad \Longrightarrow \qquad 
H(t) \equiv \frac{\dot{a}}{a} 
\quad, \quad 
\epsilon(t) \equiv - \frac{\dot{H}}{H^2} 
\; . \label{geometry}
\end{equation}
The accelerated expansion ($H > 0$ with $0 \leq \epsilon < 1$) of primordial
inflation rips virtual particles out of the vacuum \cite{Schrodinger:1939}.
The phenomenon is largest for particles such as massless, minimally coupled 
(MMC) scalars and gravitons, which are both massless and not conformally
invariant \cite{Lifshitz:1945du,Grishchuk:1974ny}. This is what is thought to 
have caused the primordial spectra of gravitons \cite{Starobinsky:1979ty} and 
scalars \cite{Mukhanov:1981xt}.

Because more and more quanta are created as inflation progresses, correlators
which involve interacting MMC scalars and gravitons often show secular growth 
in the form of powers of $\ln[a(t)]$ \cite{Onemli:2002hr,Prokopec:2002uw,
Miao:2006pn,Miao:2006gj,Kahya:2006hc,Prokopec:2008gw,Glavan:2013jca,
Wang:2014tza,Tan:2021lza,Tan:2022xpn}. For example, the 2-loop dimensionally
regulated and fully renormalized expectation value of the stress tensor of 
an MMC scalar with a $\frac{\lambda}{4!} \phi^4$ self-interaction, on de Sitter
background ($\epsilon = 0$) takes the perfect fluid form $\langle T_{\mu\nu}
\rangle = (\rho + p) u_{\mu} u_{\nu} + p g_{\mu\nu}$ with energy density
and pressure \cite{Onemli:2002hr,Onemli:2004mb,Kahya:2009sz},
\begin{eqnarray}
\rho(t) &\!\!\! = \!\!\!& 
\frac{\lambda H^4}{2^7 \pi^4} \!\times\! \ln^2(a)
+ O(\lambda^2) \; , 
\qquad \label{phi4rho} \\
p(t) &\!\!\! = \!\!\!& 
\frac{\lambda H^4}{2^7 \pi^4} \Bigl\{ 
-\ln^2(a) - \tfrac23 \ln(a) \Bigr\} + O(\lambda^2) 
\; . \qquad \label{phi4pres}
\end{eqnarray}
In the correlators of this theory each factor of $\lambda$ can be associated
with as many as two factors of $\ln(a)$. Contributions which saturate this
bound are known as {\it leading logarithm}, whereas those which have fewer
factors of $\ln(a)$ are known as {\it subleading}. In expression 
(\ref{phi4pres}) the factor of $-\ln^2(a)$ is a leading logarithm whereas 
the factor of $-\frac23 \ln(a)$ is subleading.

During a prolonged period of inflation factors of $\ln[a(t)]$ can grow so
large that they overwhelm even the smallest coupling constant. Developing a
technique to sum up the series of leading logarithms may eventually be as
important for cosmology as the renormalization group summation of leading 
momentum logarithms was to flat space quantum field theory. The late Alexei
Starobinsky solved this problem for scalar potential models 
\cite{Starobinsky:1986fx},
\begin{equation}
\mathcal{L} = 
- \frac12 \partial_{\mu} \phi \partial_{\nu} \phi g^{\mu\nu} \sqrt{-g} 
- \frac{\lambda}{4!} \phi^4 \sqrt{-g} 
\; . \label{potmodel}
\end{equation}
Starobinsky's technique can be proven to reproduce each order's leading 
logarithms \cite{Tsamis:2005hd} and, when $V(\phi)$ is bounded from below, it 
can be summed up to give the late time limits of cosmological correlators
\cite{Starobinsky:1994bd}. However, straightforward application of the 
method fails for more general theories \cite{Miao:2006pn,Prokopec:2007ak,
Miao:2008sp}.

To understand the problem with more general theories it is necessary to
distinguish between ``Active'' fields which induce large logarithms, such
as MMC scalars and gravitons, and ``Passive'' fields which do not, such 
as conformally coupled scalars, fermions, photons and even differentiated
Active fields. One can sum the leading logarithms of theories which contain 
Passive fields without derivative interactions by integrating out the 
Passive fields, assuming the Active fields are constant in space and time.
This results in a scalar potential model which can then be treated using
Starobinsky's formalism \cite{Miao:2006pn,Prokopec:2007ak}.

The large logarithms of scalar potential models derive from what DeWitt
and Brehme termed the ``tail'' part \cite{DeWitt:1960fc} of the MMC scalar 
propagator, 
\footnote{In this de Sitter background example (\ref{tail}), the tail is 
the logarithmic term. The original work on this subject was done by Hadamard
\cite{Hadamard:1923} who considered the Cauchy problem for a class of 
linear partial differential equations which include those of Riemannian
geometries. What DeWitt and Brehme did was to cast the same asymptotic form,
Hadamard's ``elementary solution'', in the geometrical language of bitensors
and biscalars. The bitensor formalism had previously been introduced by
Ruse \cite{Ruse:1931ht} and Synge \cite{Synge:1931zz}, but not applied to
Hadamard's elementary solution.}
\begin{equation}
D = 4 
\quad \Longrightarrow \quad 
i\Delta(x;x') = 
\frac1{4\pi^2} \frac1{a a' (x \!-\! x')^2} 
- \frac{H^2}{8 \pi^2} \ln\Bigl[ \frac14 H^2 (x \!-\!x')^2 \Bigr]
\; . \label{tail}
\end{equation}
The situation when derivative interactions are present is more comp\-li\-cat\-ed
because renormalization provides a new mechanism for generating large 
logarithms. When using dimensional regularization it turns out that 
$D$-de\-pen\-dent factors of $a(t)$ do not occur in divergent primitive 
contributions, which are incompletely canceled by counterterms,
\begin{eqnarray}
\lefteqn{\Bigl( {\rm Primitive:} \;\; 
\frac{(2 H)^{D-4}}{D \!-\! 4} \Bigr) -
\Bigl( {\rm Counterterm:} \;\; 
\frac{(\mu a)^{D-4}}{D \!-\! 4} \Bigr) }
\nonumber \\
& & \hspace{7.5cm}
= - \ln\Bigl( \frac{\mu a}{2 H} \Bigr) + O(D \!-\! 4) 
\; . \qquad \label{renorm}
\end{eqnarray}
These sorts of large logarithms are correctly described by a variant of
the renormalization group \cite{Miao:2021gic}.

Derivative interactions are also more complicated in the way one integrates
out Passive fields in the presence of a constant Active background. When
derivative interactions are absent the coupling to a constant Active field
is typically a mass term, as in Yukawa \cite{Miao:2006pn} or scalar quantum
electrodynamics in Lorenz gauge \cite{Prokopec:2007ak}. This leads to a
conventional Coleman-Weinberg potential \cite{Coleman:1973jx}, of course
modified by functions of the dimensionless ratio of $\frac{\phi}{H}$. 
Derivative interactions lead to a new type of effective potential in which 
the field strength of a Passive field depends upon constant Active fields 
\cite{Miao:2021gic}. In this paper we consider a third type of effective 
potential in which constant Active fields shift the Hubble parameter on 
which the Passive field propagators depend.
\footnote{Hubble-induced effective potentials can also occur in 
theories for which matter fields assume a non-trivial background 
\cite{Katuwal:2023wtl}.}

The model we shall study is how a loop of MMC scalars on de Sitter 
background induce changes in gravity,
\begin{equation}
\mathcal{L} = 
- \frac12 \partial_{\mu} \phi \partial_{\nu} \phi g^{\mu\nu} \sqrt{-g} 
+ \frac{[R \!-\! (D\!-\!2) \Lambda] \sqrt{-g}}{16 \pi G} 
\; , \label{ourmodel}
\end{equation}
where $G$ is Newton's constant and $\Lambda = (D-1) H^2$ is the 
cosmological constant. The scalar loop contribution to the graviton 
self-energy was made about a decade ago \cite{Park:2011ww}, and used to 
solve the effective field equations for gravitational radiation 
\cite{Park:2011kg,Leonard:2014zua} and for the gravitational response to 
a point mass \cite{Park:2015kua}. Unfortunately, the original computation 
fails to be conserved because it lacks a finite renormalization of the 
cosmological constant, the necessity for which was only realized recently 
\cite{Tsamis:2023fri}. This problem was compounded by representing the 
non-conserved graviton self-energy using a sum of structure functions 
which are automatically conserved \cite{Leonard:2014zua}. When the 
computation was re-done, including the finite renormalization, and making 
no assumptions about structure functions, a somewhat different result was 
found for the $\ln[a(t)]$ correction to the potentials \cite{Miao:2024atw}. 
The effects on gravitational radiation were also computed to enough 
accuracy to reveal a logarithmic change in the electric component of the 
Weyl tensor. The purpose of this paper is to explain both logarithmic 
corrections using a variant of the renormalization group. We will also 
explain how to integrate the scalar out of the gravitational field 
equation for constant graviton background, and show that the induced 
stress tensor implies precisely the finite renormalization of the 
cosmological constant which is needed to make the graviton self-energy 
conserved \cite{Tsamis:2023fri}.

This paper consists of six sections. In Section 2 we review non-linear 
sigma models, which manifest large logarithms from renormalization 
(\ref{renorm}) and also stochastic logarithms from effective potentials 
derived from Active fields modifying the field strengths of Passive 
fields. Section 3 shows that integrating the scalar out of (\ref{ourmodel}) 
in the presence of a spacetime constant graviton background induces a 
new type of effective stress tensor by modifying the Hubble constant on 
which the scalar propagator depends. We show that this result explains 
the finite renormalization required in the exact calculation. Because 
the induced stress tensor is only valid at leading logarithm order, it 
is not conserved when the graviton field is allowed to depend on space 
and time. In Section 4 we show how the stress tensor can be extended to 
give a fully conserved form. A variant of the renormalization group is
used in Section 5 to explain the large logarithms found for 
gravitational radiation and for the response to a point mass. Section 
6 gives our conclusions and also discusses the prospects for extending
this analysis to loops of gravitons.

\section{Non-linear Sigma Models}

Non-linear sigma models have the same $\, h \partial h \partial h \,$ 
derivative interactions as gravity, and induce the same factors of 
$\ln[a(t)]$ on de Sitter background, but without the complex index 
structure or the gauge issue. They have therefore received much attention 
as a simple venue for sorting out the complexities of derivative interactions 
\cite{Tsamis:2005hd,Kitamoto:2010et,Kitamoto:2011yx,Kitamoto:2018dek,
Miao:2021gic,Woodard:2023rqo,Litos:2023nvj}. A simple example consists 
of two scalar fields $A(x)$ and $B(x)$ with the following Lagrangian,
\begin{equation}
\mathcal{L} = 
-\frac12 \partial_{\mu} A \; \partial_{\nu} A \; 
g^{\mu\nu} \sqrt{-g}
- \frac12 \Bigl(1 \!+\! \frac12 \lambda A\Bigr)^2 
\partial_{\mu} B \; \partial_{\nu} B \; 
g^{\mu\nu} \sqrt{-g} 
\; . \label{LAB}
\end{equation}
The first variations of its action provide the equations 
of motion,
\begin{eqnarray}
\frac{\delta S[A,B]}{\delta A(x)} &\!\! = \!\!& 
\partial_{\mu} \Bigl[ \sqrt{-g} g^{\mu\nu} \partial_{\nu} A \Bigr] 
- \frac12 \lambda \Bigl( 1 \!+\! \frac12 \lambda A \Bigr)
\partial_{\mu} B \; \partial_{\nu} B \; g^{\mu\nu} \sqrt{-g} 
= 0
\; , \qquad \label{Avar1} \\
\frac{\delta S[A,B]}{\delta B(x)} &\!\! = \!\!& 
\partial_{\mu} \Bigl[ \Bigl (1 \!+\! \frac12 \lambda A \Bigr)^2 
\sqrt{-g} \; g^{\mu\nu} \partial_{\nu} B \Bigr] 
= 0
\; . \label{Bvar1}
\end{eqnarray}
The above theory is not renormalizable and hence requires 
an ever-increasing number of counterterms. 
\\ [5pt]
\noindent
{\bf *} {\it Perturbative Results}

The quantities computed were the expectation values 
of $A, \, B$ and their squares, the mode functions 
$u_A(\eta, k) \, \& \, u_B(\eta, k)$, and the exchange 
potentials $P_A(\eta, r) \; \& \; P_B(\eta, k)$. 
The latter two are obtained from the scalar self-mass 
$-i M^2 (x;x')$ which supplies the quantum corrections 
to the linearized effective field equation for a scalar 
$\Phi(x)$,
\begin{equation}
\mathcal{D} \Phi(x) - \int \!\! d^4x' M^2(x;x') \Phi(x') 
= J(x)
\; , \label{EFeqn}
\end{equation}
where $\, {\mathcal D} \equiv \partial^{\mu} a^2 
\partial_{\mu} \,$ is the kinetic operator and $J(x)$ 
the source. The quantum corrections to the propagation 
of scalar radiation are imprinted in the mode function 
correction which is obtained when $J(x)=0$. The choice 
$J(x) = a(\eta) \delta^3(\mathbf{x})$ gives the scalar 
exchange potential.

The above perturbative calculations of self-masses and 
VEV's were performed in the $``in-in"$ formalism which 
is causal and which allows only real self-masses 
\cite{Schwinger:1960qe,Mahanthappa:1962ex,Bakshi:1962dv,
Bakshi:1963bn,Keldysh:1964ud,Chou:1984es,Jordan:1986ug,
Calzetta:1986ey,Ford:2004wc}. The regulation technique 
is dimensional regularization which preserves coordinate 
invariance. As a result the Lagrangian (\ref{LAB}) 
requires two counterterms per scalar field to renormalize 
the self-masses at 1-loop,
\begin{eqnarray}
\Delta \mathcal{L}_{-iM^2} &\!\! = \!\!&
-\frac12 C_{\scriptscriptstyle A1} \, 
\square A \, \square A \, \sqrt{-g} 
-\frac12 C_{\scriptscriptstyle A2} \, 
R \, \partial_{\mu} A \, \partial_{\nu} A \, g^{\mu\nu} \sqrt{-g} 
\nonumber \\
& &
- \frac12 C_{\scriptscriptstyle B1} \, 
\square B \, \square B \, \sqrt{-g} 
- \frac12 C_{\scriptscriptstyle B2} \, 
R \, \partial_{\mu} B \, \partial_{\nu} B \, g^{\mu\nu} \sqrt{-g} 
\; . \label{Cterms1}
\end{eqnarray}
Moreover, the VEV's of the squares of the scalar fields
require composite operator renormalization which at 1-loop 
and 2-loop orders implies the following counterterms,
\footnote{No renormalizations are needed for the VEV's
of $A(x)$ which is ultraviolet finite to this order and
the VEV of $B(x)$ vanishes to all orders by virtue of 
the shift symmetry of (\ref{LAB}).}
\begin{eqnarray}
A_{\rm ren}^2 &\!\! = \!\!&
A^2 + K_{\scriptscriptstyle A1} R + K_{\scriptscriptstyle A2} 
R \, A^2 + K_{\scriptscriptstyle A3} R^2 + O(\lambda^4)
\; , \label{Cterms2A} \\
B_{\rm ren}^2 &\!\! = \!\!& 
B^2 + K_{\scriptscriptstyle B1} R + K_{\scriptscriptstyle B2} 
R \, B^2 + K_{\scriptscriptstyle B3} R^2 + O(\lambda^4)
\; . \label{Cterms2B}
\end{eqnarray}
Finally, to renormalize the 3-point vertex of the theory 
at 1-loop order, the counterterm Lagrangian is,
\begin{eqnarray}
\Delta \mathcal{L}_{ABB} &\!\! = \!\!&
-\frac12 C_{\scriptscriptstyle ABB1} \, \square A \, 
\partial_{\mu} B \, \partial_{\nu} B \, g^{\mu\nu} \sqrt{-g} 
-\frac12 C_{\scriptscriptstyle ABB2} \, \partial_{\mu} A \, 
\partial_{\nu} B \, \square B \, g^{\mu\nu} \sqrt{-g}
\nonumber \\
& & \hspace{-0.3cm} 
-\frac12 C_{\scriptscriptstyle ABB3} \, A \, 
\square B \, \square B \, \sqrt{-g} 
- \frac12 C_{\scriptscriptstyle ABB4} \, R \, A \, 
\partial_{\mu} B \, \partial_{\nu} B \, g^{\mu\nu} \sqrt{-g} 
\; . \qquad \label{Cterms3}
\end{eqnarray}
In (\ref{Cterms3}) the last term is the curvature-dependent
coupling constant renormalization. It turns out that at 1-loop 
there is no such renormalization and the associated 1-loop
$\beta$-function vanishes \cite{Miao:2021gic, Woodard:2023rqo},
\begin{eqnarray}
\delta\lambda &\!\! \equiv \!\!& 
\lambda_{\rm ren} - \lambda
= C_{\scriptscriptstyle ABB4} \times R + O(\lambda^5) 
= 0 \, \lambda^3 + O(\lambda^5)
\; \Rightarrow 
\label{deltalambda} \\
\beta &\!\! \equiv \!\!& 
\frac{\partial \delta\lambda}{\partial \ln(\mu)}
= 0 \, \lambda^3 + O(\lambda^5)
\;\; . \label{beta} 
\end{eqnarray}

All the coefficients in (\ref{Cterms1}-\ref{Cterms3}) are 
determined so that they absorb the primitive divergences 
of the perturbative diagrams. The leading logarithm 
renormalized results are most conveniently presented 
in the form of Table~\ref{ABresults} 
\cite{Miao:2021gic, Woodard:2023rqo}:

\begin{table}[H]
\setlength{\tabcolsep}{8pt}
\def\arraystretch{1.5}
\centering
\begin{tabular}{|@{\hskip 1mm }c@{\hskip 1mm }||c|}
\hline
Quantity & Leading Logarithms \\
\hline\hline
$u_{A}(\eta,k)$ & $\Bigl\{1 {\color{red} -\frac{\lambda^2 H^2}{32 \pi^2} 
\ln(a) } + O(\lambda^4)\Bigr\} \times \frac{H}{\sqrt{2 k^3}}$ \\
\hline
$u_{B}(\eta,k)$ & $\Bigl\{1 + 0 + O(\lambda^4)\Bigr\} \times 
\frac{H}{\sqrt{2 k^3}}$ \\
\hline
$P_{A}(\eta,r)$ & $\Bigl\{1 {\color{red} - \frac{\lambda^2 H^2}{32 \pi^2} 
\ln(a) } {\color{green} + \frac{\lambda^2 H^2}{32 \pi^2} \ln(Hr)} + 
O(\lambda^4)\Bigr\} \times \frac{KH}{4\pi} \ln(Hr)$ \\
\hline
$P_{B}(\eta,r)$ & $\Bigl\{1 {\color{green} - \frac{\lambda^2 H^2}{32 \pi^2} 
\ln(Hr) } + O(\lambda^4)\Bigr\} \times \frac{KH}{4\pi} \ln(Hr)$ \\
\hline
$\langle \Omega \vert A(x)\vert \Omega \rangle$ & $\Bigl\{1 
{\color{red} + \frac{\lambda^2 H^2}{64 \pi^2} \ln(a)} +
O(\lambda^4)\Bigr\} \times {\color{red} \frac{\lambda H^2}{16 \pi^2} 
\ln(a)}$ \\
\hline
$\langle \Omega \vert A^2(x)\vert \Omega \rangle_{\rm ren}$ & 
$\Bigl\{1 {\color{red} - \frac{\lambda^2 H^2}{64 \pi^2} \ln(a) }
+ O(\lambda^4)\Bigr\} \times {\color{red} \frac{H^2}{4\pi^2} \ln(a)}$ \\
\hline
$\langle \Omega \vert B(x)\vert \Omega \rangle$ & $0$ \\
\hline
$\langle \Omega \vert B^2(x)\vert \Omega \rangle_{\rm ren}$ & 
$\Bigl\{1 {\color{green} + \frac{3 \lambda^2 H^2}{32 \pi^2} \ln(a)}
+ O(\lambda^4) \Bigr\} \times {\color{red} \frac{H^2}{4\pi^2} \ln(a)}$ \\
\hline
\end{tabular}
\caption{\footnotesize Leading logarithm renormalized perturbative 
results for the AB model to leading logarithm. \textcolor{red}{Red}
denotes leading logarithms explained by the stochastic formalism 
while \textcolor{green}{green} denotes those explained by the 
renormalization group.}
\label{ABresults}
\end{table}

\noindent
{\bf *} {\it Re-summation Techniques: Stochastic} 

For the infrared secular contributions due to the 
ever-increasing number of degrees of freedom with
super-horizon wavelengths $\, k < a(t)H$, the stochastic 
method can re-sum the leading contributions and provide
the late time evolution; this was explicitly displayed by 
Starobinsky and Yokoyama \cite{Starobinsky:1994bd} for a 
scalar $\Phi(x)$ in de Sitter spacetime with a non-derivative 
self-interacting potential $V(\Phi)$ and a static late time 
evolution limit. The basic idea is to replace $\Phi(x)$ 
with a stochastic field $\varphi(x)$ which commutes with 
itself $[\varphi(x), \varphi(x')] = 0$, and whose correlators 
are completely free of ultraviolet divergences. The stochastic 
field $\varphi(x)$ is constructed from the same free creation 
and annihilation operators that appear in $\Phi(x)$ in such 
a way that the two fields produce the same leading logarithms 
at each order in perturbation theory. The requirement to 
concentrate on these infrared modes, simplifies the equation 
of motion to a Langevin equation obeyed by the stochastic 
field $\varphi(x)$,
\footnote{For the purposes of this analysis we have
converted from conformal time $\eta$ to co-moving time
$t$.}
\begin{equation}
{\dot \varphi}(x) - {\dot \varphi}_o(x)
= - \frac{1}{3H} V'({\varphi})
\;\; , \label{Langevin1}
\end{equation}
where $\varphi_o$ is the stochastic jitter given by
the infrared truncated free field mode sum emanating
from $\Phi(x)$. 

{\bf -} The extension of the method to scalars with derivative 
self-interactions can be seen by noting, for instance in the 
equation of motion (\ref{Avar1}) for $A(x)$, that a constant 
$A(x)={\overline A}$ field is a field strength renormalization 
of $B(x)$. It follows that we can quantify the effect of the 
undifferentiated $A(x)$ to all orders by simply integrating 
out the differentiated $B(x)$ fields in (\ref{Avar1}) for 
constant $A(x)={\overline A}$. Taking the VEV of (\ref{Avar1}) 
gives the first order equation for the 1-loop effective potential,
\begin{eqnarray}
- V'_{\rm eff}({\overline A}) \, a^D &\!\!\! = \!\!\!& 
- \frac12 \lambda \Bigl( 1 \!+\! \frac12 \lambda {\overline A} 
\Bigr) a^{D-2} 
\Bigl\langle \Omega \Bigl\vert \partial^{\mu} B(x) \, 
\partial_{\mu} B(x) \Bigr\vert \Omega \Bigr\rangle_{\overline A} 
\; , \\
&\!\!\! = \!\!\!& 
+ \frac{\frac12 \lambda (D \!-\! 1) k H^2 a^D}
{1 \!+\! \frac12 \lambda {\overline A}} 
\; , \label{Veffprime}
\end{eqnarray}
which upon taking the unregulated $D \!=\! 4$ limit and 
integrating gives the 1-loop effective potential,
\begin{equation}
V_{\rm eff}(A) = -\frac{3 H^4}{8 \pi^2} \ln\Bigl\vert 1 
+ \frac12 \lambda A \Bigr\vert 
\; . \label{Veff}
\end{equation}
By integrating the derivative interactions out of the 
field equations in the presence of a constant scalar 
background, we obtained a curvature-dependent effective 
potential $V_{\rm eff}$ for which the standard stochastic 
procedure applies: one merely replaces $V$ in the 
non-derivative re-summation rule with $V_{\rm eff}$.

{\bf -} When the generic field equation (\ref{Avar1}) 
of $A(x)$ is restricted to de Sitter spacetime we get,
\footnote{Since the scale factor varies much faster 
than the field during inflation, it is preferable 
to have derivatives act on the scale factor instead 
of the field. Thus, the single time derivative of the 
Hubble friction term dominates over the second time 
and space derivatives terms.}
\begin{equation}
-\frac{d}{dt} \Bigl( a^3 \dot{A} \Bigr) 
- V_{\rm eff}'(A) a^3 = 0 
\qquad \Longrightarrow \qquad 
\dot{A} =  \frac{\lambda H^3}{16 \pi^2}
\frac1{1 \!+\! \frac12 \lambda A} 
\; . \label{Aevolution}
\end{equation}
Equation (\ref{Aevolution}) can be solved exactly,
and for the initial condition $A_{\rm in} = 0$ the 
solution is,
\begin{equation}
A_{\rm cl} = \frac{2}{\lambda} \Biggl[ 
\sqrt{1 + \frac{\lambda^2 H^2}{16 \pi^2} \ln(a)} 
- 1 \Biggr] 
\; . \label{Acl}
\end{equation}
The Langevin equation (\ref{Langevin1}) for the 
associated stochastic field ${\mathcal A}(x)$,
\begin{equation}
\dot{\mathcal{A}} - \dot{\mathcal{A}}_o = 
\frac{\lambda H^3}{16 \pi^2}
\frac1{1 \!+\! \frac12 \lambda \mathcal{A}} 
\; , \label{stochAeqn}
\end{equation}
introduces the stochastic jitter coming from the
infrared truncated free field $\mathcal{A}_o$.
The general solution consists of the $\comp$-number 
solution (\ref{Acl}) plus a series in powers of 
${\mathcal A}_o$ coming from iterating 
(\ref{stochAeqn}),
\begin{equation}
\mathcal{A} = \mathcal{A}_{\rm cl} + \mathcal{A}_o
- \frac{\lambda^2 H^3}{32 \pi^2} \int_0^t dt' \mathcal{A}_o
+ \frac{\lambda^3 H^3}{64 \pi^2} \int_0^t dt' \mathcal{A}_o^2
+ O(\lambda^4)
\; . \label{stochsol}
\end{equation}

\noindent
{\bf *} {\it Re-summation Techniques: Renormalization Group} 

For the leading logarithms coming from the ultraviolet 
sector, the standard QFT re-summation methodology of 
the renormalization group needs, as well, an extension 
due to the presence of a curved spacetime: counterterms 
can be regarded as curvature-dependent renormalizations 
of the bare parameters present in the original theory.
Consider, for instance, the renormalization of the composite 
operator $A^2(x)$. Of the general set of counterterms 
(\ref{Cterms2A}), the $\, K_{\scriptscriptstyle A2} 
R \, A^2 \,$ is a field strength renormalization,
\begin{equation}
A^2 = \sqrt{Z_{\scriptscriptstyle A^2}} \!\times\! A_{\rm ren}^2
\quad , \quad
Z_{\scriptscriptstyle A^2} = 1 - 2 K_{\scriptscriptstyle A2} 
\!\times\! R + O(\lambda^4)
\; , \label{ZA2}
\end{equation}
which is curvature dependent.

{\bf -} The re-summation of the leading logarithms from 
the renormalization group is well known in QFT: the 1-loop
approximation to the running coupling re-sums the leading
logarithmic behaviour of all the loop diagrams.
\footnote{Furthermore, the running coupling obtained by 
solving the renormalization group equations with the 
$\beta$-function approximated with its $\ell$-loop 
expression re-sums not only the leading logarithm arising 
at any order, but also the first $\ell$-1 subleading 
logarithms.}
Recall that the $N$-point function 
$G_{N}(x_1,\dots, x_N; \lambda; \mu)$ at a scale $\mu$ can 
be expressed in terms of its value at a scale $\mu_0$ via 
the running coupling constant $\overline{\lambda}(\mu)$,
\begin{eqnarray}
G_{N}(x_1,\dots, x_N; \lambda; \mu) &\!\! = \!\!&
G_{N}(x_1,\dots, x_N; \overline{\lambda}(\mu); \mu_0)
\nonumber \\
&\mbox{}& \hspace{1.9cm} \times 
\exp \! \left[ -N \int_{\mu_0}^{\mu} \frac {d\mu'}{\mu'}
\, \gamma \big( \overline{\lambda}(\mu') \big) \right]
\; , \label{RGresum}
\end{eqnarray}
where $\gamma(\overline{\lambda}(\mu))$ is the 
$\gamma$-function of the associated field.

Since for the AB model the 1-loop $\beta$-function vanishes,
so that $\overline{\lambda}(\mu) = \lambda$, the leading 
$\ln(\mu)$ re-summation becomes a power law,
\begin{equation}
G_{N}(x_1,\dots, x_N; \lambda; \mu) =
G_{N}(x_1,\dots, x_N; \lambda; \mu_0)
\times
\left[ \frac {\mu_0}{\mu} \right]^{N \, \gamma(\lambda)}
\; . \label{RGresumAB}
\end{equation}

\noindent
{\bf *} {\it Agreement: Stochastic Analysis}

To substantiate that the above re-summations do indeed
produce the leading logarithms, we must compare their 
predictions against the perturbative results of Table~1.

{\bf -} Starting from the stochastic re-summation, we 
expand the stochastic solution (\ref{stochsol}) in powers
of $\lambda$ and take its VEV,
\footnote{The infrared truncated free field $\mathcal{A}_o$
has the canonically commuting creation and annihilation
operators of $A$ so that:
$\langle \Omega \vert \mathcal{A}_o(t,\vec{x})
\vert \Omega \rangle = 0$ and
$\langle \Omega \vert \mathcal{A}_o^2(t,\vec{x})
\vert \Omega \rangle = 
\frac{H^2 \ln(a)}{4 \pi^2}.$}
\begin{equation}
\Bigl\langle \Omega \Bigl\vert \mathcal{A}(t,\vec{x}) 
\Bigr\vert \Omega \Bigr\rangle = 
\frac{\lambda H^2}{16 \pi^2} \ln(a) \Biggl\{1 + 
\frac{\lambda^2 H^2}{64 \pi^2} \ln(a) + 
O(\lambda^4) \Biggr\} 
\; , \label{VEVstochA}
\end{equation}
as well as the VEV of its square,
\begin{equation}
\Bigl\langle \Omega \Bigl\vert \mathcal{A}^2(t,\vec{x}) 
\Bigr\vert \Omega \Bigr\rangle =
\frac{H^2}{4 \pi^2} \ln(a) \Biggl\{ 
1 - \frac{\lambda^2 H^2}{64 \pi^2} \ln(a) + 
O(\lambda^4) \Biggr\}
\; . \label{VEVstochA2}
\end{equation}
Comparing (\ref{VEVstochA}) and (\ref{VEVstochA2})
with their perturbative counterparts 
$\langle \Omega \vert A(x) \vert \Omega \rangle$ 
and
$\langle \Omega \vert A^2(x) \vert \Omega \rangle$
in Table~1 shows perfect agreement, a highly non-trivial 
achievement indeed.
\\ [6pt]
{\bf *} {\it Agreement: Renormalization Group}

A similar non-trivial agreement is maintained when
we consider the renormalization group analysis and its
re-summation of the leading logarithms. For instance, 
the Callan-Symanzik equation for the composite 
operator $A^2(x)$ has the form \cite{Miao:2021gic},
\begin{equation}
\Bigl[ a \frac{\partial}{\partial a} 
+ \beta \frac{\partial}{\partial \lambda} + 
\gamma_{\scriptscriptstyle A^2} \Bigr] 
\Bigl\langle \Omega \Bigl\vert A^2(x) 
\Bigr\vert \Omega \Bigr\rangle_{\rm ren} = 0 
\; . \label{CZeqnA2}
\end{equation}
The $\beta$ and $\gamma$ functions are,
\begin{equation}
\beta \equiv \frac{\partial \delta\lambda}{\partial \ln(\mu)}
= 0 + O(\lambda^5)
\quad , \quad
\gamma_{\scriptscriptstyle A^2} \equiv 
\frac{\partial \ln(Z_{\scriptscriptstyle A^2})}{\partial \ln(\mu^2)} 
= 0 + O(\lambda^4)
\; . \label{A2betagamma}
\end{equation}
Therefore, there should no leading logarithms coming 
from this analysis to this order (2-loop) and indeed 
the corresponding perturbative result seen in Table~1 
for $\langle \Omega \vert A^2(x) \vert \Omega 
\rangle_{\rm ren}$ confirm this; both leading logarithms 
are stochastic effects as described above.
\footnote{The same is true for the renormalized VEV 
of $A(x)$.}

{\bf -} The curvature-dependent field strength renormalization
of the composite operator $B^2(x)$ is \cite{Miao:2021gic},
\begin{equation}
Z_{\scriptscriptstyle B^2} = 
1 - 2 K_{\scriptscriptstyle B2} \!\times\! R + O(\lambda^4)
\quad \Longrightarrow \quad
\gamma_{\scriptscriptstyle B^2} = -\frac{3 \lambda^2 H^2}{16 \pi^2} 
+ O(\lambda^4 H^4)
\; , \label{B2betagamma}
\end{equation}
so that its associated Callan-Symanzik equation predicts
the leading logarithm contribution to this order to be
$+\frac{3 \lambda^2 H^2}{16 \pi^2}$ which is identical 
to what is seen in Table~1 for 
$\langle \Omega \vert B^2(x) \vert \Omega \rangle_{\rm ren}$.
\\ [9pt]
{\bf *} {\it Comments}

{\bf -} It turns out that all entries in Table~1 are 
reproduced by one {\it or} the other of the re-summation 
techniques. Yet another example, is the $\lambda^0$ order 
term $\frac{H^2}{4 \pi^2} \ln(a)$ that appears in the VEV's 
of Table~1: it comes from the coincidence limit of the 
scalar propagator and is a stochastic effect; it could 
not be explained by the renormalization group since it 
does not obey the Callan-Symanzik equation.

{\bf -} Among the physical conclusions from the above analysis,
we could highlight the behaviour of the scalar field $A(x)$.
It provides an example of a field which, {\it (i)} rolls 
down its potential at a faster rate due to its stochastic
jitter, {\it (ii)} develops a positive mass from a 
stochastically induced effective potential, and {\it (iii)}
has a time evolution that does not approach a static 
limit but grows and persists to arbitrarily late times.

\section{Constant Graviton Induced Stress Tensor}

On $D=4$ dimensional de Sitter the cosmological constant 
is $\Lambda = 3 H^2$. The gravitational field equation 
associated with our model (\ref{ourmodel}) is,
\begin{equation}
R_{\mu\nu} - \frac12 R \, g_{\mu\nu} + 3 H^2 g_{\mu\nu}
= 8 \pi G \left[ \delta^{\rho}_{~\mu} \delta^{\sigma}_{~\nu} 
- \frac12 g_{\mu\nu} g^{\rho\sigma} \right] 
\partial_{\rho} \phi \, \partial_{\sigma} \phi
\; . \label{eomD=4}
\end{equation}
We seek to integrate out the scalar fields on the right hand
side of (\ref{eomD=4}). If this could be done for an arbitrary
metric field $g_{\mu\nu}(x)$ it would give us the exact 1-loop
effective field equation, however, this is not possible. What 
we can do instead is to integrate out the scalars for constant
graviton field, which should suffice at leading logarithm order.
As we shall see, this constant background is just de Sitter 
spacetime with a different cosmological constant.
\\ [5pt]
{\bf -} Consider the general class of conformally 
rescaled backgrounds with constant $H$ and arbitrary
$\widetilde{g}_{\mu\nu}(x)$,
\begin{equation}
g_{\mu\nu}(x) \equiv 
a^2 \, \widetilde{g}_{\mu\nu}(x) \equiv 
a^2 \Big[ \eta_{\mu\nu} + \kappa h_{\mu\nu}(x) \Big] 
\quad , \quad 
a = -\frac1{H \eta}
\; . \label{background}
\end{equation}
Here and henceforth $\kappa^2 \equiv 16 \pi G$ is the loop-counting
parameter of quantum gravity. For geometries (\ref{background}) we
find,
\footnote{We define:
$\widetilde{\Gamma}^{\rho}_{~\mu\nu} \equiv 
\frac12 \widetilde{g}^{\rho\sigma} \Bigl(
\widetilde{g}_{\sigma\mu , \nu} 
+ \widetilde{g}_{\nu\sigma , \mu} 
- \widetilde{g}_{\mu\nu , \sigma} \Bigr)$.}
\begin{eqnarray}
& & \hspace{-1.9cm}  
\Gamma^{\rho}_{~\mu\nu} = 
a H \Bigl( \delta^{\rho}_{~\mu} \delta^{0}_{~\nu}
+ \delta^{\rho}_{~\nu} \delta^{0}_{~\mu} 
- \widetilde{g}^{0\rho} \, \widetilde{g}_{\mu\nu} \Bigr) 
+ \widetilde{\Gamma}^{\rho}_{~\mu\nu}
\; , \label{Gamma} \\   
& & \hspace{-1.9cm}  
R^{\rho}_{~\sigma\mu\nu} =
-H^2 \, \widetilde{g}^{00} \Bigl( 
\delta^{\rho}_{~\mu} \, g_{\sigma \nu} 
- \delta^{\rho}_{~\nu} \, g_{\sigma\mu} \Bigr)
\nonumber \\
& & \hspace{-0.5cm} 
+ \, a H \Bigl[ 
\delta^{\rho}_{~\mu} \widetilde{\Gamma}^{0}_{~\nu\sigma}
- \delta^{\rho}_{~\nu} \widetilde{\Gamma}^{0}_{~\mu\sigma} 
+ \widetilde{g}^{\rho \alpha} \Bigl(
\widetilde{\Gamma}^{0}_{~\mu\alpha} \, \widetilde{g}_{\nu\sigma} 
- \widetilde{\Gamma}^{0}_{~\nu\alpha} \, \widetilde{g}_{\mu\sigma}
\Bigr) \Bigr] 
+ \widetilde{R}^{\rho}_{~\sigma\mu\nu}
\; , \label{Riemann} \\
& & \hspace{-1.9cm}  
R_{\mu\nu} = 
- 3H^2 \, \widetilde{g}^{00} g_{\mu\nu}
+ a H \Bigl[ 
2 \widetilde{\Gamma}^{0}_{~\mu\nu} 
+ \widetilde{g}_{\mu\nu} \, \widetilde{g}^{\alpha\beta} \,
\widetilde{\Gamma}^{0}_{~\alpha\beta}\Bigr] 
+ \widetilde{R}_{\mu\nu}
\; , \label{RicciT} \\
& & \hspace{-1.9cm} 
R =
- 12 H^2 \, \widetilde{g}^{00}
+ 6 \frac{H}{a} \, \widetilde{g}^{\alpha\beta} \,
\widetilde{\Gamma}^{0}_{~\alpha\beta} 
+ \frac1{a^2} \, \widetilde{R}
\; . \label{RicciS} 
\end{eqnarray}
Thus, the left hand side of (\ref{eomD=4}) becomes,
\begin{eqnarray}
& & \hspace{-1.9cm} 
R_{\mu\nu} - \frac12 R \, g_{\mu\nu} + 3H^2 g_{\mu\nu} 
= 
3H^2 \Bigl( 1 + \widetilde{g}^{00} \Bigr) g_{\mu\nu}
+ 2 a H \Bigl[ \widetilde{\Gamma}^{0}_{~\mu\nu} 
\!-\! \widetilde{g}_{\mu\nu} \, \widetilde{g}^{\alpha\beta} \,
\widetilde{\Gamma}^{0}_{~\alpha\beta} \Bigl] 
\nonumber \\
& & \hspace{2.7cm}
+ \widetilde{R}_{\mu\nu} 
- \frac12 \widetilde{g}_{\mu\nu} \widetilde{R}
\; . \label{eomLHS}
\end{eqnarray}
\noindent
{\bf -} When we focus on geometries with $h_{\mu\nu}(x)$
constant (which implies constant $\widetilde{g}_{\mu\nu}(x)$),
so that only terms without derivatives acting on the field
are selected, the curvature tensor (\ref{Riemann}) simplifies, 
\begin{equation}
R^{\rho}_{~\sigma\mu\nu} \Big\vert_{\widetilde{g}_{\mu\nu} = c}
=
-H^2 \, \widetilde{g}^{00} \Bigl( 
\delta^{\rho}_{~\mu} \, g_{\sigma \nu} 
- \delta^{\rho}_{~\nu} \, g_{\sigma\mu} \Bigr)
\; , \label{RiemannConst}
\end{equation}
and we recognize a de Sitter geometry albeit with a different 
cosmological constant,
\begin{equation}
\widetilde{g}_{\mu\nu , \rho} = 0 
\quad \Longrightarrow \quad 
H^2 
\; \longrightarrow \;
-\widetilde{g}^{00} H^2 
\; . \label{replacement}
\end{equation}

\noindent
{\bf -} To integrate out the differentiated scalars present 
in the right hand side of (\ref{eomD=4}),
\begin{equation}
\partial_{\mu} \phi(x) \, \partial_{\nu} \phi(x)
\;\; \Longrightarrow \;\;
\Bigl\langle \Omega \, \Bigl\vert
\partial_{\mu} \phi(x) \, \partial_{\nu} \phi(x)
\Bigr\vert \, \Omega \Bigr\rangle
\; , \label{VEV}
\end{equation}
the above expectation value is to be evaluated in the
scalar vacuum for the background (\ref{background})
with $\widetilde{g}_{\mu\nu}(x)$ a spacetime constant.

What is needed for the computation of (\ref{VEV}) is 
the coincidence limit of the mixed derivative of the 
scalar propagator in de Sitter spacetime, a quantity 
that has been evaluated using dimensional regularization
(and the old Hubble parameter $H$) \cite{Onemli:2002hr,
Onemli:2004mb},
\begin{eqnarray}
& & \hspace{-2.9cm} 
\lim_{x' \to x} 
\partial_{\mu} \partial'_{\nu} \, i\Delta_{dS}(x;x') = 
- \frac{H^D}{(4\pi)^{\frac{D}2}} \, 
\frac{\Gamma(D)}{2 \Gamma(\frac{D}2 \!+\!1)} 
\times g_{\mu\nu}(x) \Big\vert_{\widetilde{g}_{\mu\nu} = c} 
\nonumber \\
& & \hspace{0.8cm}
\longrightarrow \;
-\frac{3 H^4}{32 \pi^2} 
\times g_{\mu\nu}(x) \Big\vert_{\widetilde{g}_{\mu\nu} = c} 
\; . \label{dd'propcoinc}
\end{eqnarray}
The finite $D \!=\! 4$ remainder in (\ref{dd'propcoinc}) 
emerges after the automatic subtraction of the quartic 
and quadratic divergences of the doubly differentiated 
propagator. 

By effecting the replacement (\ref{replacement}) in
the evaluation (\ref{dd'propcoinc}), the right hand
side (\ref{VEV}) equals,
\begin{equation}
\Bigl\langle \Omega \, \Bigl\vert 
\partial_{\mu} \phi(x) \, \partial_{\nu} \phi(x)
\Bigr\vert \; \Omega 
\Bigr\rangle_{g_{\mu\nu} = a^2 \widetilde{g}_{\mu\nu}} 
=
-\frac{3 \, [-\widetilde{g}^{00} H^2]^2}{32 \pi^2} 
\!\times\! g_{\mu\nu}(x) 
\; , \label{newVEV}
\end{equation}
so that,
\begin{eqnarray}
& & \hspace{-1.8cm}
8 \pi G \, \Bigl\langle \Omega \, \Bigl\vert 
\partial_{\mu} \phi \, \partial_{\nu} \phi
- \frac12 g_{\mu\nu} g^{\rho\sigma} 
\partial_{\rho} \phi \, \partial_{\sigma} \phi
\Bigr\vert \, \Omega 
\Bigr\rangle_{g_{\mu\nu} = a^2 \widetilde{g}_{\mu\nu}} 
\nonumber \\
& & \hspace{5.9cm}
= \frac{3 \kappa^2 [-\widetilde{g}^{00} H^2]^2}{64 \pi^2} 
\times g_{\mu\nu}(x)
\, . \label{eomRHS}
\end{eqnarray}
Note that (\ref{eomRHS}) represents a negative contribution 
to the cosmological constant which can become arbitrarily 
large if $\, -\widetilde{g}^{00}$ achieves arbitrarily large
values.
\\ [5pt]
{\bf -} Substituting (\ref{eomLHS}) and (\ref{eomRHS}) into 
the gravitational field equations (\ref{eomD=4}) we arrive 
at the leading logarithm form,
\begin{eqnarray}
& & \hspace{-2.1cm}
\widetilde{R}_{\mu\nu} 
- \frac12 \widetilde{g}_{\mu\nu} \widetilde{R}
+ 2 a H \Bigl[ 
\widetilde{\Gamma}^{0}_{~\mu\nu}
\!-\! \widetilde{g}_{\mu\nu} \, \widetilde{g}^{\alpha\beta} 
\, \widetilde{\Gamma}^{0}_{~\alpha\beta} \Bigl]
\nonumber \\
& & \hspace{2.9cm}
= - 3 H^2 \left[ 1 + \widetilde{g}^{00} 
- \frac{\kappa^2 H^2}{64 \pi^2} (\widetilde{g}^{00})^2 
\right] g_{\mu\nu}  
\; , \label{EFTeom}
\end{eqnarray}
displaying the scalar-induced gravitational stress-energy
tensor $T^{\rm ind}_{\mu\nu} \,$,
\begin{equation}
T^{\rm ind}_{\mu\nu} = 
\frac{3}{32 \pi^2} 
\Big[ - \widetilde{g}^{00} H^2 \Big]^2 g_{\mu\nu}
\; , \label{Tmnind}
\end{equation}
always with the understanding that 
$\, \widetilde{g}_{\mu\nu}(x) \!=\! {\rm constant}$.

For $\widetilde{g}^{00} \!=\! -1$ the induced stress tensor 
(\ref{Tmnind}) takes the form of a negative addition to the 
cosmological constant of,
\begin{equation}
\Delta \Lambda = 
- \frac{3 \kappa^2 H^4}{64 \pi^2} 
\; . \label{cosmoshift}
\end{equation}
If we want the quantity ``$H$'' to represent the true Hubble 
parameter, at least initially, (\ref{cosmoshift}) must be 
absorbed by making a finite renormalization of the cosmological 
constant. This is precisely the finite renormalization which 
was previously recognized as being necessary to make the scalar 
loop contribution to the graviton self-energy conserved --- 
see equations (33) and (38) of \cite{Tsamis:2023fri}. 

\section{Generally Conserved Extension}

The induced stress tensor (\ref{Tmnind}) is not conserved when 
$\widetilde{g}_{\mu\nu}$ is allowed to depend on space and time. 
It is important to understand that this is not an error but rather 
the inevitable consequence of working at leading logarithm order. 
Consider, for example, the $\lambda \phi^4$ energy density and 
pressure (\ref{phi4rho}-\ref{phi4pres}). Under the symmetries 
of homogeneity and isotropy conservation reduces to $\dot{\rho} 
= -3 H (\rho + p)$, and relations (\ref{phi4rho}-\ref{phi4pres}) 
do obey this,
\begin{eqnarray}
\dot{\rho} &\!\!\! = \!\!\!& 
\frac{\lambda H^4}{2^7 \pi^2} \times 
2 H \ln(a) + O(\lambda^2) 
\; , \label{LHS} \\
-3 H (\rho + p) &\!\!\! = \!\!\!& 
-3 H \Biggl\{ \frac{\lambda H^4}{2^7 \pi^2}
\Bigl[ + \ln^2(a) - \ln^2(a) - \tfrac23 \ln(a) \Bigr] 
+ O(\lambda^2) \Biggr\}
\; . \qquad \label{RHS}
\end{eqnarray}
However, the right hand side (\ref{RHS}) only agrees with the left 
(\ref{LHS}) by virtue of the sub-leading factor of $-\frac23 \ln(a)$ 
in the pressure (\ref{phi4pres}). With just the $\, \pm \ln^2(a)$ 
leading logarithm contributions, conservation is violated.

Although the violation of conservation of the induced stress tensor 
(\ref{Tmnind}) has a good explanation, it does present us with a problem.
The gravitational field equations (\ref{eomD=4}) consist of 10 relations,
for a general metric, of which 4 are automatically implied by conservation. 
If one adds the non-conserved stress tensor (\ref{Tmnind}) this will no
longer be true. All 10 of the equations cannot be used, because the 0th 
order equations are still conserved. So which 6 of the 10 equations should 
we solve, and why? Note that this sort of issue could not arise with the 
very similar effective potentials which are induced in non-linear sigma 
models; it is specific to gravity and it will have to be confronted as 
well when graviton loops are considered.

We believe that the answer is to extend the leading logarithm stress
tensor (\ref{Tmnind}) to a generally conserved form. The variation of any
invariant would give such a conserved form, and it is worthwhile first 
considering the two 1-loop counterterms,
\begin{equation}
\Delta \mathcal{L} = c_1 R^2 \sqrt{-g} + c_2 C_{\alpha\beta\gamma\delta}
C^{\alpha\beta\gamma\delta} \sqrt{-g} \; . \label{1loop}
\end{equation}
The variation of the Weyl counterterm will still contain a Weyl tensor,
which vanishes for de Sitter, so this counterterm has no impact. The 
variation of the $R^2$ counterterm contributes to the stress tensor as,
\begin{equation}
\Delta T^{1}_{\mu\nu} = -\frac{2}{\sqrt{-g}} \frac{\delta \Delta S_1}{\delta 
g^{\mu\nu}} = c_1 \Bigl[ g_{\mu\nu} R^2 - 4 \Bigl( R_{\mu\nu} + g_{\mu\nu}
\square - D_{\mu} D_{\nu}\Bigr) R\Bigr] \; . \label{DS1}
\end{equation}
For $D$-dimensional de Sitter with arbitrary Hubble parameter 
$\overline{H}$, the Ricci tensor is $R_{\mu\nu} = (D-1) \overline{H}^2 
g_{\mu\nu}$. The covariant derivatives all vanish so we have,
\begin{equation}
g_{\mu\nu} R^2 - 4 \Bigl( R_{\mu\nu} + g_{\mu\nu} \square - 
D_{\mu} D_{\nu}\Bigr) R = D (D\!-\!1)^2 (D\!-\!4) \overline{H}^4 
g_{\mu\nu} \; . \label{deSitterDS1}
\end{equation}
The factor of $(D-4)$ means that only the divergent part of the 
counterterm $c_1$ can make a nonzero contribution to expression 
(\ref{DS1}),
\begin{equation}
c_1 \longrightarrow \frac{\mu^{D-4}}{D\!-\!4} \!\times\! \frac1{2^7 
\!\cdot\! 3^3 \!\cdot\! \pi^2} 
\end{equation}
The result does not quite agree with (\ref{Tmnind}),
\begin{equation}
\Delta T^{1}_{\mu\nu} = \frac{\overline{H}^4}{32 \pi^2} g_{\mu\nu}
\; . \label{T1mn}
\end{equation}

Although the $R^2$ counterterm induces a stress tensor (\ref{T1mn}) 
which fails to agree with the leading logarithm result (\ref{Tmnind}),
it is close enough that we are motivated to consider more general
functions of the Ricci scalar. Expressions (\ref{RicciT}) and 
(\ref{RicciS}) for the Ricci tensor and scalar suggest that we could 
regard $[- \widetilde{g}^{00} H^2]$ as $\frac{R}{12}$ and try to 
extend the stress-energy tensor (\ref{Tmnind}) so that it emerges 
from a more general Lagrangian,
\begin{equation}
\mathcal{L}_T = -f(R) \sqrt{-g}
\; , \label{LT}
\end{equation}
which gives the conserved source,
\begin{equation}
-\frac{\kappa^2}{\sqrt{-g}} 
\frac{\delta S_T[g]}{\delta g^{\mu\nu}(x)} = 
- \frac12 \kappa^2 \, g_{\mu\nu} f(R) 
+ \kappa^2 \Bigl[ R_{\mu\nu} + g_{\mu\nu} \square 
- D_{\mu} D_{\nu} \Bigr] f'(R) 
\; , \label{sourceLT}
\end{equation}
where $\, \square \equiv \frac{1}{\sqrt{-g}} \partial_{\mu}
[ \sqrt{-g} g^{\mu\nu} \partial_{\nu} ]$ is the scalar
covariant d'Alembertian and $D_{\mu}$ is the covariant
derivative operator. By substituting (\ref{RicciT}) and
(\ref{RicciS}) into (\ref{sourceLT}), by discarding the 
sub-leading terms involving $\widetilde{\Gamma}^0_{~\alpha\beta}$
and $\widetilde{R}$, and by defining $X \equiv -12 \, 
\widetilde{g}^{00} H^2$, equation (\ref{sourceLT}) reduces 
to a 1st order differential equation for the function 
$f(X)$,
\begin{equation}
\frac{X^2}{2^9 \cdot 3 \cdot \pi^2}
= - f(X) + \frac12 X f'(X)
= \frac{X^3}{2} \Big( \frac{f(X)}{X^2} \Big)'
\; , \label{feqn}
\end{equation}
which has the solution,
\begin{equation}
f(X) = \frac{X^2 \ln(\frac{X}{12 H^2})}{2^8 \cdot 3 \cdot \pi^2}
\; , \label{fsol}
\end{equation}
whose 1st and 2nd derivatives equal,
\begin {equation}
f'(X) = \frac{2X \ln(\frac{X}{12 H^2}) + X}{2^8 \cdot 3 \cdot \pi^2}
\quad , \quad 
f''(X) = \frac{2 \ln(\frac{X}{12 H^2}) + 3}{2^8 \cdot 3 \cdot \pi^2}
\; . \label{f'f''}
\end{equation}

\noindent
{\bf -} Furthermore, the two derivative terms present 
in (\ref{sourceLT}) are,
\begin{eqnarray}
g_{\mu\nu} \square &\!\!\! = \!\!\!& 
\widetilde{g}_{\mu\nu} \Bigl[
\widetilde{g}^{\rho\sigma} \partial_{\rho} \partial_{\sigma} 
+ 2 a H \, \widetilde{g}^{0\rho} \partial_{\rho} 
- \widetilde{g}^{\rho\sigma} \widetilde{\Gamma}^{\alpha}_{~\rho\sigma} 
\partial_{\alpha} \Bigr] 
\; , \label{Box} \\
- D_{\mu} D_{\nu} &\!\!\! = \!\!\!& 
- \partial_{\mu} \partial_{\nu} 
+ a H \Bigl[ \delta^0_{~\mu} \partial_{\nu} 
+ \delta^0_{~\nu} \partial_{\mu} 
- \widetilde{g}_{\mu\nu} \widetilde{g}^{0\rho} \partial_{\rho} \Bigr] 
+ \widetilde{\Gamma}^{\rho}_{~\mu\nu} \partial_{\rho} 
\; . \label{DmDn}
\end{eqnarray}
It will turn out to be rather useful to extract the 
metric from the Ricci tensor (\ref{RicciT}),
\begin{eqnarray}
R_{\mu\nu} &\!\!\! = \!\!\!& 
\frac14 g_{\mu\nu} R 
+ 2 a H \Bigl[ \widetilde{\Gamma}^{0}_{~\mu\nu} 
\!-\! \frac14 \widetilde{g}_{\mu\nu} \, 
\widetilde{g}^{\alpha\beta} \,
\widetilde{\Gamma}^{0}_{~\alpha\beta}\Bigr] 
+ \Bigl[ \widetilde{R}_{\mu\nu} 
\!-\! \frac14 \widetilde{g}_{\mu\nu} \widetilde{R} \Bigr] 
\nonumber \\
&\!\!\! \equiv \!\!\!&
\frac14 g_{\mu\nu} R + \Delta R_{\mu\nu}
\; , \label{DeltaRicciT}
\end{eqnarray}
and to define the dimensionless Ricci scalar $\mathcal{R}$
from (\ref{RicciS}),
\begin{equation}
R =
12 H^2 \Big[ \! - \widetilde{g}^{00}
+ \frac{\widetilde{g}^{\alpha\beta} \,
\widetilde{\Gamma}^{0}_{~\alpha\beta}}{2 a H} 
+ \frac{\widetilde{R}}{12 a^2 H^2} \Big]
\equiv
12 H^2 \times {\mathcal{R}}
\; . \label{calRicciS} 
\end{equation}
In view of (\ref{DeltaRicciT}), (\ref{calRicciS}) and
(\ref{fsol}), (\ref{f'f''}) the three terms on the right
hand side of (\ref{sourceLT}) become,
\begin{eqnarray}
& & 
-\frac12 \kappa^2 \, g_{\mu\nu} \, f(R) 
\; = \;
-6 \epsilon H^2 \mathcal{R}^2
\ln(\mathcal{R}) \, g_{\mu\nu} 
\; , \label{sourceA} \\
& &
\kappa^2 R_{\mu\nu} f'(R) 
\; = \;
3 \epsilon H^2 \mathcal{R}^2 \Bigl[
2 \ln(\mathcal{R}) + 1 \Bigr] g_{\mu\nu} 
+ \epsilon \mathcal{R} \Bigl[
2 \ln(\mathcal{R}) + 1 \Bigr] 
\Delta R_{\mu\nu} 
\; , \qquad \label{sourceB} \\
& &
\kappa^2 \Bigl[
g_{\mu\nu} \square \!-\! D_{\mu} D_{\nu} \Bigr] f'(R) 
\; = \; 
\epsilon \Bigl[ 
g_{\mu\nu} \square \!-\! D_{\mu} D_{\nu} \Bigr] 
\Bigl[ 2 \mathcal{R} \ln(\mathcal{R}) + \mathcal{R} \Bigr] 
\; , \label{sourceC}
\end{eqnarray}
where $\epsilon \equiv \frac{\kappa^2 H^2}{64 \pi^2}$.

Summing (\ref{sourceA}-\ref{sourceC}) gives,
\begin{eqnarray}
& & \hspace{-0.9cm}
\frac{\kappa^2}{\sqrt{-g}} 
\frac{\delta S_T[g]}{\delta g^{\mu\nu}(x)} 
= 
3 \epsilon H^2 \mathcal{R}^2 \, g_{\mu\nu}
+ \epsilon \mathcal{R} \Bigl[2 \ln(\mathcal{R}) + 
1\Bigr] \Delta R_{\mu\nu} 
\nonumber \\
& & \hspace{3.3cm}
+ \epsilon \Bigl[ 
g_{\mu\nu} \square - D_{\mu} D_{\nu} \Bigr] 
\Bigl[ 2 \mathcal{R} \ln(\mathcal{R}) + \mathcal{R} \Bigr] 
\; . \qquad \label{newsource}
\end{eqnarray}

The spacetime geometries of cosmological interest 
are homogeneous,
\begin{equation}
\widetilde{g}_{00} = -\frac1{x(t)} 
\qquad , \qquad 
\widetilde{g}_{0i} = 0 
\qquad , \qquad 
\widetilde{g}_{ij} = \delta_{ij} 
\; . \label{homobackground}
\end{equation}
The quantities of interest have the following form,
\begin{eqnarray}
& &
\widetilde{\Gamma}^{0}_{~00} = -\frac{a \dot{x}}{2 x} 
\qquad \Longrightarrow \qquad 
\widetilde{g}^{\alpha\beta} \widetilde{\Gamma}^{0}_{~\alpha\beta}
= \frac12 a \dot{x} 
\; , \label{homoconnection} \\
& &
\widetilde{R}_{\mu\nu} = 0
\qquad , \qquad
\Delta R_{\mu\nu} = 
- a^2 H \dot{x} \Bigl[ \, 
\frac1{x} \delta^{0}_{~\mu} \delta^{0}_{~\nu} 
+ \frac14 \widetilde{g}_{\mu\nu} \Bigr] 
\; , \qquad \label{homoRicciT} \\
& &
\mathcal{R} = x + \frac{\dot{x}}{4 H}
\; , \label{homoRicciS}
\end{eqnarray}
so that the relevant derivative operators become,
\begin{equation}
g_{\mu\nu} \square - D_{\mu} D_{\nu} = 
-g_{\mu\nu} \Bigl[ x \partial_t + 2 x H 
+ \frac12 \dot{x} \Bigr] \partial_t 
- \frac{a^2}{x} \delta^{0}_{~\mu} \delta^{0}_{~\nu} \Bigl[ 
x \partial_t - x H + \frac12 \dot{x} \Bigr]
\partial_t 
\; . \label{homoDDs}
\end{equation}
Consequently, the two non-zero components in this 
background are,
\begin{eqnarray}
\frac{\kappa^2}{\sqrt{-g}} \frac{\delta S_T[g]}{\delta g^{00}(x)} 
&\!\!\! = \!\!\!& 
\epsilon \, g_{00} \times A
\; , \label{rhoT} \\
& & \hspace{-2.5cm}
A \; \equiv \; 
3 H^2 \mathcal{R}^2 
+ 3 H \Bigl( \frac14 \dot{x} - x \partial_t \Bigr) 
\Bigl[2 \mathcal{R} \ln(\mathcal{R}) + \mathcal{R} \Bigr] 
\; . \label{A} \\
\frac{\kappa^2}{\sqrt{-g}} \frac{\delta S_T[g]}{\delta g^{ij}(x)} 
&\!\!\! = \!\!\!& 
\epsilon \, g_{ij} \times B
\; , \label{pT} \\
& & \hspace{-2.5cm}
B \; \equiv \; 
3 H^2 \mathcal{R}^2
- \Bigl( \frac14 H \dot{x} + 2 H x \partial_t 
+ \frac12 \dot{x} \partial_t + x \partial_t^2 \Bigr) 
\Bigl[ 2 \mathcal{R} \ln(\mathcal{R}) + \mathcal{R} \Bigr] 
\; , \qquad \label{B}
\end{eqnarray}
and the conservation equation is satisfied,
\begin{equation}
\dot{A} = 3 H (-A + B) 
\; . \label{conservationT}
\end{equation}
Moreover, the extended source (\ref{newsource}) has 
the proper correspondence limit (\ref{Tmnind}) because
when $\widetilde{g}_{\mu\nu}$ is constant only the
first term in the right hand side of (\ref{newsource})
survives. It should also be noted that the extension
we analyzed is by no means unique.

\section{Renormalization Group Analysis}

Analysis of the exact 1-loop effective field equations 
produced three results \cite{Miao:2024atw}. First, the electric 
component of the Weyl tensor for plane wave gravitational 
radiation ($h_{ij}(t,\vec{x}) = \epsilon_{ij}(\vec{k},\lambda) 
u(t,k) e^{i \vec{k} \cdot \vec{x}}$) receives a logarithmic 
correction,
\begin{equation}
C_{0i0j}(t,\vec{x}) = 
C^{(0)}_{0i0j}(t,\vec{x}) \Bigl\{
1 - \tfrac{3 G H^2}{10 \pi} \!\times\! \ln(a) 
+ O(G^2)\Bigr\} 
\; . \label{Weyl}
\end{equation}
A similar logarithmic correction affects the gravitational
potentials,
\begin{equation}
ds^2 = 
- \Bigl[1 \!-\! 2 \Psi(t,r)\Bigr] dt^2 
+ a^2(t) \Bigl[ 1 - 2 \Phi(t,r) \Bigr] 
d\vec{x} \!\cdot\! d\vec{x} 
\; , \label{PsiPhi}
\end{equation}
generated in response to a static point mass $M$,
\begin{eqnarray}
\Psi(t,r) &\!\!\! = \!\!\!& 
\tfrac{G M}{a r} \Bigl\{ 
1 + \tfrac{G}{20 \pi a^2 r^2} - \tfrac{3 G H^2}{10 \pi} 
\!\times\! \ln(a H r) + O(G^2) \Bigr\} 
\; , \label{Psi} \\
\Phi(t,r) &\!\!\! = \!\!\!& 
- \tfrac{G M}{a r} \Bigl\{ 
1 - \tfrac{G}{60 \pi a^2 r^2} - \tfrac{3 G H^2}{10 \pi} 
\!\times\! \Bigl[ \ln(a H r) \!+\! 1 \Bigr] + O(G^2) \Bigr\} 
\; . \label{Phi}
\end{eqnarray}
The factors of $\frac{G}{a^2 r^2}$ in (\ref{Psi}-\ref{Phi}) 
represent de Sitter versions of corrections which have long 
been known on flat space background \cite{Capper:1973bk,
Donoghue:1993eb,Donoghue:1994dn}. They are not leading 
logarithm contributions and have no interest for us.

The three leading logarithm corrections (\ref{Weyl}) and 
(\ref{Psi}-\ref{Phi}) are not explained by the induced stress
tensor we considered in the previous section. They closely 
resemble the logarithmic corrections previously found to the
electric components of the field strength tensor for plane
wave electromagnetic radiation \cite{Wang:2014tza}, and to the
Coulomb potential for a static point charge \cite{Glavan:2013jca}.
Those two results can be explained by a variant of the 
renormalization group \cite{Glavan:2023tet}, and it turns out 
that a similar explanation works for the gravitational logarithms
(\ref{Weyl}) and (\ref{Psi}-\ref{Phi}).

The model (\ref{ourmodel}) was the first quantum gravitational 
system ever studied using dimensional regularization 
\cite{tHooft:1974toh}. It has long been known that this model
requires two counterterms at 1-loop order,
\begin{equation}
\Delta \mathcal{L} = 
c_1 R^2 \sqrt{-g} 
+ c_2 C_{\alpha\beta\gamma\delta} C^{\alpha\beta\gamma\delta} 
\sqrt{-g} 
\; . \label{DeltaL}
\end{equation}
The coefficients $c_1$ and $c_2$ which were used to derive 
the 1-loop results (\ref{Weyl}) and (\ref{Psi}-\ref{Phi}) 
are \cite{Park:2011ww,Miao:2024atw},
\begin{eqnarray}
c_1 &\!\!\! = \!\!\!& 
\frac{\mu^{D-4} \Gamma(\frac{D}2)}{2^8 \pi^{\frac{D}2}} 
\frac{(D\!-\!2)}{(D\!-\!1)^2 (D\!-\!3) (D\!-\!4)}
\; , \label{c1} \\
c_2 &\!\!\! = \!\!\!& 
\frac{\mu^{D-4} \Gamma(\frac{D}2)}{2^8 \pi^{\frac{D}2}} 
\frac{2}{(D\!+\!1) (D\!-\!1) (D\!-\!3)^2 (D\!-\!4)}
\; , \label{c2}
\end{eqnarray}
where $\mu$ is the scale of dimensional regularization.

Like the non-linear sigma models considered in Section 3, 
our model (\ref{ourmodel}) is nonrenormalizable. Recall from 
Section 3 that it was necessary to sort the resulting BPHZ 
(Bogoliubov, Parasiuk \cite{Bogoliubov:1957gp}, Hepp 
\cite{Hepp:1966eg} and Zimmermann \cite{Zimmermann:1968mu,
Zimmermann:1969jj} counterterms into those which can be 
viewed as renormalizing parameters of the bare theory and 
those that provide irrelevant higher derivative contributions 
which do not induce large logarithms. Experience with the 
large electromagnetic logarithms induced by graviton loops 
\cite{Glavan:2023tet} shows that this sometimes requires 
dissecting the original counterterms. For example, 
renormalizing the 1-loop graviton contribution to the vacuum 
polarization on de Sitter background requires three counterterms 
\cite{Leonard:2013xsa,Glavan:2015ura},
\footnote{The non-covariant $\Delta C$ counterterm arises 
from the unavoidable breaking of de Sitter invariance in 
the graviton propagator.}
\begin{eqnarray}
\lefteqn{\Delta \mathcal{L} = 
\Delta C \, H^2 F_{ij} F_{k\ell} g^{ik} g^{j\ell} \sqrt{-g} 
+ \overline{C} \, H^2 F_{\mu\nu} F_{\rho\sigma} 
g^{\mu\rho} g^{\nu\sigma} \sqrt{-g} } 
\nonumber \\
& & \hspace{0.7cm}
+ \, C_4 \, D_{\alpha} F_{\mu\nu} \, D_{\beta} F_{\rho\sigma} \,
g^{\alpha\beta} g^{\mu\rho} g^{\nu\sigma} \sqrt{-g} 
\; . \qquad \qquad \label{EMcterms}
\end{eqnarray}
However, the curvature-dependent field strength renormalization 
which explains the large electromagnetic logarithms is 
\cite{Glavan:2023tet},
\begin{equation}
\delta Z = -4 \Bigl[ \overline{C} - (3 D \!-\! 8) C_4\Bigr] H^2 \; .
\label{EMdeltaZ}
\end{equation}

The analogous decomposition of the gravitational counterterms
(\ref{DeltaL}) begins by writing the Eddington ($R^2$) counterterm 
as the sum of three terms,
\begin{equation}
R^2 = \Bigl( R \!-\! D \Lambda\Bigr)^2 
+ 2 D \Lambda \Bigl[R \!-\! (D\!-\!2) \Lambda \Bigr] 
+ D (D\!-\! 4) \Lambda^2 
\; . \label{Eddington}
\end{equation}
{\it (i)} The first term in (\ref{Eddington}) involves products 
of second derivatives of the graviton field, and explicit 
computation shows that it engenders no large logarithms 
\cite{Miao:2024atw}. \\
{\it (ii)} However, the second term can be viewed either as 
a renormalization of Newton's constant in the original model 
(\ref{ourmodel}), or else as a renormalization of the graviton 
field strength. Because we have no information about higher order 
couplings it is convenient to adopt the latter interpretation. \\
{\it (iii)} The third term is finite. 

The interpretation adopted for the second term in (\ref{Eddington})
applies as well to the special term arising, at quadratic order, 
from the term proportional to $\, \partial^2_0 h_{\nu\sigma}$ 
in the Weyl counterterm of (\ref{DeltaL}),
\begin{equation}
C_{\mu\nu\rho\sigma} C^{\mu\nu\rho\sigma} = 
- \tfrac{2\kappa}{a^4}
\, \partial_{\mu} \partial_{\rho} h_{\nu\sigma} 
\, \widetilde{C}^{\mu\nu\rho\sigma} + O(\kappa^3 h^3) 
\; \longrightarrow \; 
- \tfrac{2 \kappa}{a^4} \, \partial_0^2 h_{ij} 
\, \widetilde{C}_{0i0j} 
\; , \label{Weylcterm}
\end{equation}
where $\widetilde{C}^{\mu\nu\rho\sigma}$ is the conformally
rescaled Weyl tensor.
Direct calculation \cite{Miao:2024atw} shows that the two time 
derivatives can act on scale factors to eventually contribute 
to the large logarithms (\ref{Weyl}) and (\ref{Psi}-\ref{Phi}). 

Hence the total field strength renormalization is,
\begin{equation}
\delta Z = 
D \Bigl[ 2 (D\!-\!1) c_1 - c_2 \Bigr] \kappa^2 H^2 
\quad \Longrightarrow \quad 
\gamma \equiv \frac{\partial \ln(1 \!+\! \delta Z)}
{\partial \ln(\mu^2)} 
= \frac{3 G H^2}{20 \pi} 
\; . \label{gamma}
\end{equation}

The Callan-Symanzik equation for a 2-point Green's function 
reads,
\begin{equation}
\Bigl[ \frac{\partial}{\partial \ln(\mu)} 
+ \beta_G \, \frac{\partial}{\partial G} + 2 \gamma \Bigr] 
G^{(2)} = 0 
\; . \label{CSeqn}
\end{equation}
The $\beta$-function for $G$ vanishes at the order we are working.
From expression (\ref{renorm}) we see that factors of the renormalization
scale $\mu$ come in the form $\ln(\mu a)$, so that we can replace the 
derivative with respect to $\ln(\mu)$ by a derivative with respect to
$\ln(a)$. The Weyl tensor (\ref{Weyl}) can be considered as the 2-point
correlator between the operator $C_{0i0j}$ and the single particle 
creation operator; similarly, the potentials (\ref{Psi}-\ref{Phi}) 
can be considered as 2-point correlators. It follows that the three 
large logarithms in (\ref{Weyl}) and (\ref{Psi}-\ref{Phi}) can all be 
explained using the renormalization group. If we ignore possible running
of Newton's constant it is even possible to sum up the effects to 
conclude,
\begin{eqnarray}
C_{0i0j}(t,\vec{x}) &\!\! \longrightarrow \!\!& 
C^{(0)}_{0i0j}(t,\vec{x}) \!\times\! 
\Bigl[ a(t) \Bigr]^{-\frac{3 G H^2}{10 \pi}} 
\; , \label{resumWeyl} \\
\Psi(t,r) &\!\! \longrightarrow \!\!& 
\frac{GM}{a(t) r} \!\times\! 
\Bigl[ a(t) H r \Bigr]^{-\frac{3 G H^2}{10 \pi}} 
\; , \label{resumPsi} \\
\Phi(t,r) &\!\! \longrightarrow \!\!& 
-\frac{GM}{a(t) r} \!\times\! 
\Bigl[ a(t) H r \Bigr]^{-\frac{3 G H^2}{10 \pi}} 
\; . \label{resumPhi}
\end{eqnarray}

\section{Epilogue}

The continuous production of MMC scalars and gravitons during inflation
causes loop corrections which would be constant in flat space to acquire 
secular growth factors in the form of powers of the logarithm of the scale
factor \cite{Onemli:2002hr,Prokopec:2002uw,Miao:2006pn,Miao:2006gj,
Kahya:2006hc,Prokopec:2008gw,Glavan:2013jca,Wang:2014tza,Tan:2021lza,
Tan:2022xpn}. Over a prolonged period of inflation these factors must 
overwhelm even the smallest coupling constant. At this point perturbation 
theory breaks down and one must employ some non-perturbative re-summation. 
Starobinsky's stochastic formalism \cite{Starobinsky:1986fx,
Starobinsky:1994bd} sums the series of leading logarithms of scalar potential 
models (\ref{potmodel}) but it must be extended to correctly describe the 
leading logarithms of more general models. In these models one must 
distinguish between Active fields which engender secular growth factors 
and Passive fields which do not. One form of extending Starobinsky's 
formalism is to integrate out Passive fields in the presence of a 
spacetime constant Active field background. The result is an effective 
potential which can then be treated using Starobinsky's formalism. With 
the work done in this paper we now know of three ways in which these 
effective potentials can be induced:
\begin{itemize}
\item{Through a constant Active field giving rise to a Passive 
field mass, as in Yukawa \cite{Miao:2006pn};}
\item{Through a constant Active field changing a Passive field 
strength, as in nonlinear sigma models \cite{Miao:2021gic}; and}
\item{Through a constant Active field changing the Hubble parameter 
on which a Passive field propagator depends, as occurs in our model
(\ref{ourmodel}).}
\end{itemize}

In Section 3 we considered an MMC scalar coupled to gravity
(\ref{ourmodel}) and integrated the scalar out of the gravitational
field equation in the presence of a constant graviton background. 
It turns out that this is equivalent to merely changing the de 
Sitter Hubble constant according to the rule (\ref{replacement}). The
resulting stress tensor (\ref{Tmnind}) precisely explains the finite
renormalization of the cosmological constant which was previously 
noted as being necessary to make the graviton self-energy conserved
\cite{Tsamis:2023fri}.

In Section 4 we noted that the induced stress tensor (\ref{Tmnind}) is
not conserved when one goes beyond leading logarithm order to permit
to graviton field to vary in space and time. This is not an error, but
rather the inevitable consequence of working at leading logarithm order.
However, it does pose a problem in solving the gravitational field 
equations. We showed that (\ref{Tmnind}) can be extended to a form 
(\ref{newsource}) which is generally conserved.

For theories which possess derivative interactions, such as non-linear 
sigma models and gravity, a second extension must be made of Starobinsky's
formalism to include the large logarithms induced by the incomplete 
cancellation between primitive divergences and counterterms (\ref{renorm}).
These logarithms are described by a variant of the renormalization group 
in which some portion of the BPHZ counterterms can be regarded as 
renormalizing couplings or field strengths of the original theory. 
That procedure was described in Section 2 for non-linear sigma models, 
and we applied it in Section 5 to the large logarithms (\ref{Weyl}) 
and (\ref{Psi}-\ref{Phi}) which occur in our model (\ref{ourmodel}). 
The two BPHZ counterterms (\ref{DeltaL}) can be regarded as giving 
rise to a graviton field strength renormalization (\ref{gamma}) which 
explains all three of the large logarithms. If we ignore possible 
running of Newton's constant, it is even possible to give fully resummed 
results (\ref{resumWeyl}-\ref{resumPhi}). We remark that it should 
be possible to determine any running of $G$ by computing the 1PI 
3-point function, the same as has recently been done for non-linear 
sigma models \cite{Woodard:2023rqo}.

The point of making this study was to facilitate the development of a 
leading logarithm re-summation technique for graviton loops, so it is 
worth commenting on what will generalize and what may differ. First, 
the procedure of integrating out differentiated graviton fields for 
constant $\widetilde{g}_{\mu\nu}$ should involve the same replacement 
(\ref{replacement}) as for our scalar model (\ref{ourmodel}). However, 
the more complicated form of the graviton propagator \cite{Tsamis:1992xa,
Woodard:2004ut} makes it likely that the induced stress tensor will contain
a term proportional to $\delta^0_{\mu} \delta^0_{\nu}$ in addition to one 
proportional to $g_{\mu\nu}$. Like the scalar-induced stress tensor 
(\ref{Tmnind}), it is inevitable that the graviton-induced stress tensor
will not be conserved when one permits the graviton field to depend on space
and time. We plan to seek a fully conserved extension, but we do not expect
that a local one such as (\ref{newsource}) can be found. We also expect that
1-loop corrections to the graviton mode function and to the response to a
point mass will involve stochastic fluctuations in these extended, effective
field equations, in addition to renormalization group effects. Finally, we
expect the need for more counterterms than just the two covariant 
possibilities (\ref{DeltaL}), owing to the same de Sitter breaking which 
produced the noncovariant $\Delta C$ counterterm (\ref{EMcterms}) for 
electromagnetism.

\vspace{.5cm}

\centerline{\bf Acknowledgements}

This work was partially supported by Taiwan NSTC grants 112-2112-M-006-017 
and 113-2112-M-006-013, by NSF grant PHY-2207514 and by the Institute for 
Fundamental Theory at the University of Florida.

\end{document}